\documentclass[prl,twocolumn,superscriptaddress]{revtex4}

\usepackage{graphics}

\begin{document}

\title{Comment on ``Anomalous proximity effect in underdoped 
{YBa\(_2\)Cu\(_3\)O\(_{6+x}\)} Josephson junctions''}

\author{J. Quintanilla}
\affiliation{ Departamento de F\'{\i}sica e Inform\'atica, 
Instituto de F\'{\i}sica de S\~ao Carlos, Universidade de S\~ao Paulo, 
Caixa Postal 369, 13560-970 S\~ao Carlos, S\~ao Paulo, Brazil}

\author{K. Capelle}
\affiliation{Departamento de Qu\'{\i}mica e F\'{\i}sica Molecular, 
Instituto de Qu\'{\i}mica de S\~ao Carlos, Universidade de S\~ao Paulo, 
Caixa Postal 780, 13560-970 S\~ao Carlos, S\~ao Paulo, Brazil}
\author{L.N. Oliveira}

\affiliation{ Departamento de F\'{\i}sica e Inform\'atica, 
Instituto de F\'{\i}sica de S\~ao Carlos, Universidade de S\~ao Paulo, 
Caixa Postal 369, 13560-970 S\~ao Carlos, S\~ao Paulo, Brazil}
\date{\today}

\pacs{pacs numbers}

\maketitle

\newcommand{\be}{\begin{equation}}
\newcommand{\ee}{\end{equation}}
\newcommand{\bi}{\bibitem}

Recently Decca et al. have reported an unusually long-ranged proximity effect
(PE)
between the superconducting and insulating phases of YBCO \cite{decca}.  These
beautiful experiments have received lots of attention because  they may help to
discriminate between alternative scenarios for cuprate
superconductivity \cite{decca,scenarios}. However, later work  \cite{scenarios}
took for granted the claim \cite{decca} that such long range
is anomalous. Here we point out that this is not warranted.

\begin{figure}

{\centering
\resizebox*{0.78\columnwidth}{!}{\includegraphics{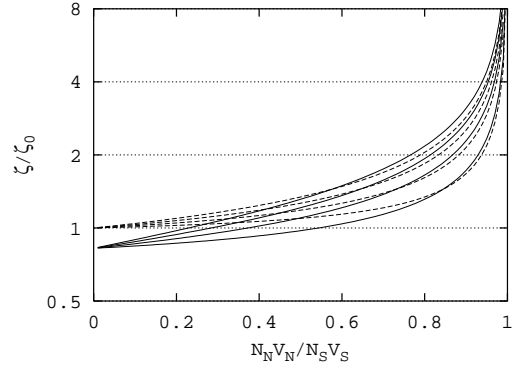}} 
\par}

\caption{\label{figure}Dependence of the range of the proximity effect on
the parameters of the interface. Solid lines: results obtained from
Eq.~(\ref{less_wrong}), valid for $N_S=N_N\equiv N$. 
Dashed lines: results obtained
from Eq.~(4.15b) of \protect\cite{degennes}, valid also for
$N_S \neq N_N$. Each line corresponds to a different value of $N_S V_S =
0.1,0.2,0.3,0.4$ (from bottom to top).}
\end{figure}

At the heart of our argument is the formula used in \cite{decca} to 
estimate the expected range of the PE,
$\zeta\approx\zeta_0\equiv \sqrt{ {\hbar D}/{2\pi k_{B}T} }$,
where $D$ is the diffusion constant. Using this expression \cite{decca} 
obtains $\zeta = 15\,\mbox{nm}$, which is indeed much smaller than the experimental 
value $\zeta^{exp} = 90\,\mbox{nm}$. 
However, the same formula agrees
with earlier experiments on YBCO \cite{exper}, which poses the 
question why it fails to account for the data of \cite{decca}.
To answer we first note that $\zeta\approx\zeta_0$ is only the 
$V_N \to 0$ limit of the more general expression \cite{deutscher}
\be
\frac{\zeta}{\zeta_0}=
\int _{0}^{\infty }\frac{d\eta }{\pi\eta}
\left[ -\frac{\Sigma'}{\Sigma }
+\frac{NV_{N}\Sigma'}{1-V_{N}/V_{S}+NV_{N}\Sigma }+\frac{2}{\eta }\right]
\label{less_wrong}
\ee
for an interface of a superconductor
and a normal conductor with coupling constants $V_S$ and
$V_N$, respectively. $\Sigma$ is a function defined in \cite{deutscher} and $N$ 
is the density of states at the Fermi level.
A numerical evaluation of Eq.~(\ref{less_wrong}) is presented in
Fig.~\ref{figure}.
The key point is that in the experiments of \cite{decca} superconductivity is 
induced in part of an underdoped sample by photodoping.
The PE thus takes place not between two different materials (as 
in standard experiments), or in an inhomogeneous material a region of which has been 
severely damaged in the creation of the normal region (as in \cite{exper}), but between two differently doped parts {\it of 
one and the same material}. The limit $V_N \to V_S$ is then more appropriate 
than $V_N\to 0$. In this limit Eq.~(\ref{less_wrong}) does predict $\zeta \gg 
\zeta_0$, suggesting that the long range is 
not an unconventional feature of YBCO.

At $V_N=V_S$ Eq.~(\ref{less_wrong}) predicts 
$\zeta \to \infty$. This infinity reflects the fact that in Eq.~(\ref{less_wrong})
the density of states is assumed to be the same on both sides of the interface 
($N_N=N_S=N$), so that for $V_N=V_S$ both materials are identical. 
The restriction to $N_N=N_S$ can be overcome in the
one-frequency approximation \cite{degennes}. Numerical evaluation of
Eq.~(4.15b) of \cite{degennes} yields the dashed curves in Fig.~\ref{figure}.
Clearly, a large, but finite value of \(\zeta/\zeta_0\) is obtained, e.g., if 
$V_N\approx V_S$
 and $N_S \agt N_N$ \cite{foot}.

In summary, although there are unconventional features in the experiments, the
claim that the long range of the PE is one of them \cite{decca}
seems presently unfounded. Conventional theory \cite{deutscher,degennes}
predicts comparably long ranges, once the novel nature of the experiments is
accounted for. We thus recommend performing similar experiments on conventional
superconductors. The lines of thought on cuprate
superconductivity stemming from  the assumption that $\zeta^{exp}$ is
anomalously large \cite{decca,scenarios} must also be critically reexamined.

{\it This work was funded by FAPESP and CNPq.}

\end{document}